\documentclass{phb-proc4-auth}

\usepackage{graphicx}
\usepackage{amssymb}

\begin{document}
\begin{frontmatter}

\journal{SCES '04}

\title{Quadrupolar interactions in heavy fermion metal YbRh$_2$Si$_2$}

\author[1,2,3]{R. J. Radwanski}
\author[2]{Z. Ropka}

\address[1]{Institute of Physics, Pedagogical University, 30-084 Krakow, Poland}
\address[2]{Center of Solid State Physics, S$nt$ Filip 5, 31-150 Krakow, Poland}

\corauth[3]{Corresponding Author: E-mail: sfradwan@cyf-kr.edu.pl,
http://css-physics.edu.pl}

\vspace{-0.7cm}
\begin{abstract}

We describe the experimentally revealed by Sichelschmidt et al,
Phys. Rev. Lett. 91 (2003) 156401, $g$ tensor, $g_{\bot}$=3.561
and $g_{\|}$=0.17, at 5 K by means of crystal field interactions
of the $4f^{13}$ configuration of the Yb$^{3+}$ ion of
YbRh$_2$Si$_2$ in a slightly orthorhombically distorted tetragonal
crystal field. We have shown that the temperature dependence of
the quadrupolar interactions Q(T) of the Yb nucleous will help to
distinguish between $\Gamma_7$ and $\Gamma_6$ ground state. For
the $\Gamma_7$ ground state Q(T) is expected to exhibits an
anomalous dependence. \vspace{-0.5cm}

\end{abstract}



\end{frontmatter}
Recently Sichelschmidt \textit{et al.} \cite{1} reported the first
successful Electron Spin Resonance (ESR) studies on single
crystalline heavy fermion metallic compound YbRh$_{2}$Si$_{2}$
that allowed for the unambiguous observation of the localized $f$
states in a Kondo compound. The importance of this experiment
relies in the fact, that practically all theories devoted to
heavy-fermion phenomena were taking the itinerant or band behavior
of $f$ electron as the starting point. Theoretical approaches to
heavy-fermion phenomena with localized $f$ states were simply
rejected both by referees of prestigious physical journals and by
their editors that simply prohibited any discussion of the
localized magnetism and crystal-field (CEF) interactions. Thus,
the discovery by the Prof. F. Steglich's group is really of the
great importance for theoretical understanding of heavy-fermion
compounds. The authors of Ref. \cite{1} are fully aware of the
theoretical importance of their observation putting a lot of
attention to evidence that this ESR signal comes out from the bulk
unscreened $Yb^{3+}$ moments \textit{below} Kondo temperature
$T_{K}$. By means of the ESR experiment, at the external field
B=0.188 T with the X-band frequency of 9.4 GHz, Sichelschmidt
\textit{et al.} \cite{1} revealed the existence of a localized
doublet characterized by very anisotropic $g$ tensor, with
$g_{\bot }$ = 3.561 and $g_{\Vert }$ = 0.17 at 5 K.

The aim of this paper is to report our search for the atomic-scale
CEF parameters describing the observed $g$ tensor in order to find
the microscopic origin of this localized state and its
experimental verification.

We have attributed the observed state to the Yb$^{3+}$
configuration, more exactly to the strongly-correlated 4$f^{13}$
configuration that, being in the Hund's rule ground multiplet
$^{2}F_{7/2}$, is described by $J$ = 7/2. Our studies show that
the higher multiplet $^{2}F_{5/2}$ does not affect the
ground-multiplet properties. The crystal field of the tetragonal
symmetry:

$H_{CF}^{tetr} $ = $B_{2}^{0}O_{2}^{0}$ + $B_{4}^{0}O_{4}^{0} $ + $%
B_{6}^{0}O_{6}^{0}$ + $B_{4}^{4}O_{4}^{4}$ + $B_{6}^{4}O_{6}^{4}$

splits the 8-fold degenerated multiplet $^{2}F_{7/2}$ into 4
Kramers doublets, 2$\Gamma _{6}$ and 2$\Gamma _{7}$. The lower in
energy states take the form (notation by $\sin \alpha$ and $\cos
\alpha$ assures the automatic normalization and the sign $\pm$
corresponds to 2 Kramers conjugate states):

$\Gamma _{6}^{1}$ = $\sin \alpha |\pm 1/2>$ + $\cos \alpha |\mp
7/2>$ and:

$\Gamma _{7}^{1}$ = $\sin \beta |\pm 3/2>$ + $\cos \beta |\mp
5/2>$.
We intend to reproduce the $g$ tensor keeping the overall CEF splitting $%
\Delta _{CEF}$ of size of 600-750 K (55-70 meV) and the first
excited level at $D$ of 70-100 K (6-9 meV) in order to assure the
proper thermodynamics and a reasonable magnitude of CEF
interactions.

We have found that for the \textbf{$\Gamma _{7}^{1}$ ground
state} the perfect reproduction of the ESR results, $g_{\bot }$
= 3.561 ($J_{\bot }$ = $\pm 1.56$) and $g_{\Vert }$ = 0.17 ($%
J_{\Vert }$ = $\pm 0.08$), is obtained for parameters: $B_{2}^{0}$= +14 K, $%
B_{4}^{0}$ = +60 mK, $B_{6}^{0}$ = -0.5 mK, $B_{4}^{4}$ = -2.30 K and $%
B_{6}^{4}$=-10 mK with a small local orthorhombic distortion
$B_{2}^{2}$ = +0.22 K \cite{2}. These parameters yield the ground
Kramers doublet:

$\Gamma _{7}^{1}$ = 0.803 $|\pm 3/2>$ + 0.595 $|\mp 5/2>$ - 0.027
$|\mp 1/2>$ - 0.009 $|\pm 7/2>$

that is characterized by $J_{\bot }$ = $\pm 1.560$, $J_{\Vert }$ =
$\pm 0.081 $ and $Q_{f}$ = -4.7. The excited states are at 85 K
($\Gamma _{6}^{1}$
with $Q_{f}$ = -13.8), 485 K ($\Gamma _{7}^{2}$) and 688 K ($\Gamma _{6}^{2}$%
). The admixture of last two small terms is the effect of the
local orthorhombic distortion. It is important to realize that
whatever lower symmetry is only 4 Kramers doublets always occur
for the 4$f^{13}$ configuration.

For the \textbf{$\Gamma _{6}^{1}$ ground state}, attained for
instance by tetragonal CEF
interactions $B_{2}^{0}$ = +10 K, $B_{4}^{0}$ = -75 mK, $B_{6}^{0}$ = +7.5 mK, $%
B_{4}^{4}$ = -2.22 K, $B_{6}^{4}$ = - 6.2 mK with $B_{2}^{2}$ =
+0.60 K, the eigenfunction:

$\Gamma _{6}^{1}$ = 0.944 $|\pm 1/2>$ + 0.322 $|\mp 7/2>$ - 0.052
$|\mp 3/2>$ - 0.047 $|\pm 5/2>$

yields $J_{\bot }$ =$\pm 1.561$, $J_{\Vert }$ = $\pm 0.084$ and
$Q_{f}$ of -11.2 For these parameters the excited states are at 92
K ($\Gamma _{7}^{1}$ with $Q_{f}$ = -3.4), 538 K ($\Gamma
_{6}^{2}$) and 546 K ($\Gamma _{7}^{2}$).

Both sets equally well reproduce the g tensor and quite well the
overall temperature dependence of the paramagnetic susceptibility
$\chi (T)$ and its huge anisotropy, presented in Fig. 1a of Ref.
\cite{3}, the preference for the magnetic ordering with moments
perpendicular to the c axis, the magnetization curve for external
magnetic fields up to 60 T applied along the tetragonal c axis
(Fig. 1b of Ref. \cite{3} - the magnetization at 2 K and at 60 T
amounts to 0.85 $\mu _{B}$). Inelastic-neutron-scattering results
are not known yet so it is hardly possible to distinguish these
states. We propose that they can be distinguished by the Mossbauer
experiments, in particular by measurement of the quadrupolar
moment $Q_{f}$ of the 4f shell. As is shown in Fig. 1 the
temperature dependence of Q(T) is completely different for both
states. In case of the $\Gamma _{7}^{1}$ ground state $Q_{f}$
shows a non-monotonic dependence. This anomalous dependence was
discussed in Ref. \cite{4} within the CEF-based model.
\begin{figure}[ht]
\centering
\includegraphics[width = 6.0cm]{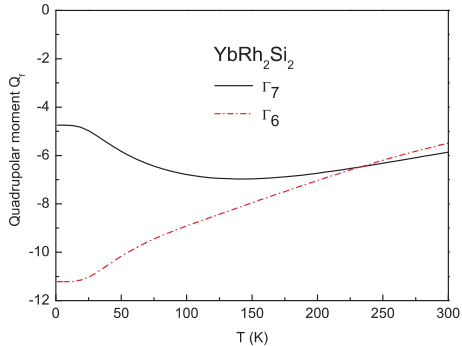} \vspace{-0.3cm}\caption{Temperature
dependence of the quadrupolar moment of the 4f shell in
YbRh$_2$Si$_2$ calculated for two different ground states
$\Gamma_6$ and $\Gamma_7$.}
\end{figure}

We do not think that the present sets of CEF parameters are the
final ones. There is a great number of sets that produce the shown
ground-state eigenfunction (the simplest can be obtained by
multiplication of all parameters by a constant positive value what
causes equal spreading of the electronic structure) but surely the
obtained sets substantially confine the searching area for CEF
parameters. Though we worked hard by last 20 years in the
evaluation of CEF and exchange interactions in different
rare-earth compounds, both ionic and intermetallics \cite{5,6} but
the existence of so well-defined (so extremely thin) CEF states in
a metallic compound YbRh$_{2}$Si$_{2}$ is a big surprise indeed.
This compound was regarded as one of the prominent heavy-fermion
compound with itinerant $f$ electrons and with a substantial Kondo
temperature (at least 25 K \cite{1}). In the developed by us
Quantum Atomistic Solid State (QUASST) Theory \cite{7,8}, we
recognize that the standard CEF approach is a giantly correlated
electron approach to compounds containing open-shell
transition-metal atoms. In our understanding the Kondo temperature
is related to the energy of the first excited CEF doublet and the
Kondo resonance is related to the removal of the Kramers
degeneracy that is a source of low-energy, below 0.2 meV,
excitations. These excitations are neutral, spin-like excitations
and they are responsible for large low-temperature specific heat,
a hallmark of heavy-fermion physics. Recently CEF states have been
also revealed in another heavy-fermion metal UPd$_{2}$Al$_{3}$,
that exhibits antiferromagnetism below 14 K and superconductivity
below 2 K. These states are related to the 5$f^{3}$ configuration
\cite{6}.

In $\textbf{conclusion}$, we have derived CEF parameters of the
tetragonal symmetry with a small orthorhombic distortion that
perfectly reproduce the ESR values ($g_{\bot }$ = 3.561 and
$g_{\Vert }$ = 0.17) as well as provide good reproduction of
thermodynamical properties both for the $\Gamma _{6}^{1}$ and
$\Gamma _{7}^{1}$ ground state. The proposed Mossbauer experiment
for the evaluation of temperature dependence of the quadrupolar
moment of the 4f shell can distinguish between these two ground
states.

\end{document}